\documentclass[reprint]{revtex4-2}

\usepackage{physics}
\usepackage{amsmath}
\usepackage{mathrsfs}
\usepackage{siunitx}
\usepackage{array}
\usepackage{graphicx}% Include figure files
\usepackage{bm}% bold math

\begin{document}

\preprint{APS/12,3-QED}

\title{Low-energy $S$-wave scattering of $\text{H}+e^-$ by a Lagrange-mesh method}
\author{Jean Servais}\email{jean.servais@ulb.be (corresponding author)}
\author{Jérémy Dohet-Eraly}\email{jdoheter@ulb.ac.be}
\affiliation{Physique Quantique, and\\
Physique Nucléaire Théorique et Physique Mathématique,\\
C.P. 229, Université libre de Bruxelles (ULB),\\
B-1050 Brussels, Belgium.\\}
\date{\today}
\begin{abstract}
A method combining the Lagrange-mesh and the complex Kohn variational methods is developed for computing the $\mathcal{S}$ matrix of a 2\,$+$\,1 elastic scattering in the frame of three-body Coulomb systems. Resonance parameters can be obtained from values of the $\mathcal{S}$ matrix at several scattering energies. The method is illustrated with the $S$-wave low-energy scattering of an electron onto hydrogen. The computed phase shifts are at least as accurate as the literature results, and the resonance parameters are more accurate than the best literature results by several orders of magnitude. Both the infinite and finite proton mass cases are considered.
\end{abstract}
\maketitle
\section{\label{intro}Introduction}

This work introduces a method for evaluating the low-energy scattering properties in three-body systems. By combining the Lagrange-mesh method in perimetric coordinates \cite{HB99,HB01,HB03,Ba15} with the Kohn variational method in its complex form~\cite{MJ87_CKVM, Lucchese89, Kievsy97}, the $\mathcal{S}$ matrix of the elastic scattering of a particle onto a two-body quantum system is obtained. This complex matrix leads directly to the phase shift, and its evaluation at several scattering energies also enables to describe resonances. Indeed, as resonances correspond to poles of the $\mathcal{S}$ matrix, these poles can be determined
by extrapolating the $\mathcal{S}$ matrix to the complex plane~\cite{RA07}. 

The present approach is illustrated by considering the low-energy $S$-wave scattering of an electron onto a hydrogen atom, between the $n_1=1$ and $n_1=2$ ionization thresholds. This system has been studied since decades until recently, and providing highly accurate values of the elastic phase shifts and resonance parameters at low scattering energies is still a scientific challenge (see Refs.~\cite{Schwartz61, Armstead68, Heller74,KH96, BT01_CKVM,CH07,ZMV08,DGH23} and references therein). \\
\indent The complex Kohn variational method was developed in 1987 by Miller and Jansen op de Haar~\cite{MJ87_CKVM} as an alternative approach to the standard real Kohn variational principle~\cite{Kohn48}. It leads directly to the complex $\mathcal{S}$ matrix rather than to the real $\mathcal{K}$ matrix. It presents the important advantage of eliminating spurious singularities usually present in the real form of this principle \cite{MJ87_CKVM,AMO23_Drake}. It is particularly relevant in the presence of Coulomb interactions, and is applied in electron-molecule as well as nuclear scattering calculations (see, for instance, Refs.~\cite{Kievsy97,BT01_CKVM}).\\ \indent 
The Lagrange-mesh method was introduced by Baye and Heenen in 1986 \cite{BH86}. It is a pseudovariational method presenting the advantage of being simple, fast, and often as accurate as a purely variational method. Almost twenty-five years ago, the Lagrange-mesh method in perimetric coordinates has proven to be successful at computing the bound states of several bielectronic three-body Coulomb systems, such as He, H$^-$, Ps$^-$, or H$_2^+$ \cite{HB99,HB01,HB03}. More recently, by combining the Lagrange-mesh method and the complex scaling method~\cite{Ho83}, highly accurate values of the resonance parameters (energies and widths) of the $S$ and $P$ states of He and Ps$^-$ have been obtained \cite{DES22,SDE23}. In addition, the resonance parameters of exotic atoms, namely He$^+\Bar{p}$ and He$^+\pi^-$, have also been obtained by this approach, for several low values of the total angular momentum $L$ \cite{BDES19,BDE21}. However, it has failed to give satisfactory results for high values of the angular momentum, in the relevant region where the capture of the exotic particle occurs (see Refs.~\cite{BDES19,BDE20,BDE21} and references therein for more information).\\
\indent The present work is a natural extension of the Lagrange-mesh method in perimetric coordinates. The combination with the complex Kohn variational principle leads to the elastic phase shifts, which are not directly accessible by means of our complex scaling approach \cite{DES22,SDE23}. This work also paves the way for evaluating accurately resonance parameters in  exotic Coulomb atoms for relatively high values of the angular momentum, as in Ref.~\cite{KS97}, for instance. \\
\indent The theoretical background is presented in Sec.~\ref{theory}. After a short description of the applied methodology, Sec.~\ref{results} reports the computed phase shifts, as well as several resonance energies and widths, for the singlet and triplet $S$ states of the hydrogen ion. A comparison with results obtained by  applying the complex scaling method is provided, following our previous work described in Ref.~\cite{DES22}, as well as with results from the literature. Both the infinite and finite proton mass cases are considered. Sec.~\ref{conclu} gives concluding remarks and perspectives.
Atomic units, in which $\hbar = e =  m_e = a_0 = 1$, are used throughout the paper. All calculations have been performed in double precision arithmetic on a standard workstation.

%--------------------------------------------------------------------
% THEORY
%--------------------------------------------------------------------

\section{\label{theory}Theory}
\subsection{Kohn variational principle}
The non-relativistic internal Hamiltonian describing a three-body quantum system under Coulomb forces is 
\begin{eqnarray}
    H =& -\dfrac{1}{2 m_1} \Delta_{\bm{r}_1}-\dfrac{1}{2 m_2} \Delta_{\bm{r}_2}-\dfrac{1}{2 m_{3}} \Delta_{\bm{r}_3}\nonumber \\&+ \dfrac{Z_1 Z_2}{r_{12}}+ \dfrac{Z_1Z_3}{r_{13}} +\dfrac{Z_2Z_3}{r_{23}}-T_{\rm{c.m.}},   \label{hamiltonian}
\end{eqnarray}
where the vectors $\bm{r}_1, \bm{r}_2$, and $\bm{r}_{3}$ refer to the positions of the three particles, $m_1$, $m_2$ and $m_3$ are the masses of the three particles, $T_{\rm{c.m.}}$ is the center-of-mass kinetic energy, $Z_1$, $Z_2$, $Z_3$ are the charges of the particles, and $r_{12}$, $r_{13}$, and $r_{23}$ are the interparticle distances.\\ \indent
In the present work, we study the $S$-wave elastic phase shifts of a bielectronic three-body system ($Z_2=Z_3=-1$, $m_2=m_3=1$) for singlet ($\sigma=0$) and triplet ($\sigma=1$) states, between the $n_1=1$ and the $n_1=2$ ionization threshold energies
\begin{equation}
    E_{{n_1}}=-\dfrac{Z_1 ^2}{2 n_1^2}\mu_{12},
\end{equation}
where $\mu_{12}$ is the reduced mass between particles 1 and 2.
Although the formulation of the Kohn variational principle is general (many-body systems involving open and closed channels), we choose to restrict the theoretical background to the single-channel case, which is relevant here.\\ \indent
The spatial wavefunction $\Psi$ of the system describing the scattering of an electron onto a two-body system formed by particle 1 and the other electron can be written as a square-integrable contribution $\Phi$ complemented with asymptotic functions $\Omega_1$ and $\Omega_2$ \cite{Kievsy97},
\begin{equation}
    {\Psi} = {\Phi}+\Omega_{1}(\bm{x}_1,\bm{x}_2)+{\Bar{\mathcal{S}}}(k)\Omega_2(\bm{x}_1,\bm{x}_2)
    \label{TotWF}
\end{equation}
where ${\Bar{\mathcal{S}}}(k)$ is an estimate of the $\mathcal{S}$ matrix. We shall see in the end of this section that an improved estimation of the $\mathcal{S}$ matrix, which we note $\mathcal{S}(k)$, is provided by the complex Kohn variational principle. The vectors $\bm{x}_1$ and $\bm{x}_2$ correspond to the Jacobi coordinates defined as
\begin{align}
    \bm{x}_1 &= \bm{r}_1-\bm{r}_2,\\
    \bm{x}_2 &= (1-\alpha)\bm{r}_1+\alpha \bm{r}_2-\bm{r}_3,
\end{align}
with $\alpha=m_2/(m_1+m_2)$. The angular momenta associated to the Jacobi coordinates are noted $l_1$ and $l_2$. As we consider $S$ states with energies comprised between the two first ionization thresholds, $l_1=l_2=0$. The element of volume in Jacobi coordinates reads, after integrating over the angles,
\begin{equation}
    \text{d}V = 16 \pi^2 x_1^2 x_2^2 \text{d}x_1 \text{d}x_2.
\end{equation}
The total wavefunction must be antisymmetric with respect to the exchange of the two electrons. Indeed, there exist a spatially symmetric singlet state ($\sigma=0$) and a spatially antisymmetric triplet state ($\sigma=1$). By applying the operator 
$[1+(-1)^{\sigma}P_{23}]/2$, where $P_{23}$ has the effect of exchanging spatially particles 2 and 3, i.e.\ the electrons, the spatial wavefunction can be properly antisymmetrized. The spatial wavefunction of Eq.~\eqref{TotWF} is thus replaced by 
\begin{equation}
    {\Psi}^{\sigma}\equiv[ {\Psi}+(-1)^{\sigma}  \Tilde{\Psi}]/2,
\end{equation}
with $\Tilde{\Psi}=P_{23}\Psi$. The square-integrable part of $\Psi^{\sigma}$ is expressed as the linear combination of $N_t$ basis functions,
\begin{equation}
    {\Phi}^{\sigma}=\sum_{l=1}^{N_t} C_l \phi_l^{\sigma},
    \label{square-integrable-part}
\end{equation}
which are defined in Sec.~\ref{LMMsection}. The asymptotic wavefunctions read, for $\lambda=1,2$
(see Ref.~\cite{Kievsy97}),
\begin{eqnarray}
    \Omega_{\lambda}^{\sigma}(\bm{x}_1,\bm{x}_2)&&=\dfrac{1+(-1)^{\sigma}P_{23}}{8 \pi}\sqrt{{2 k \mu_{12,3}}} \dfrac{R_{10} (x_1)}{k x_2}\nonumber \\
    \times [i F_{0}&&(k x_2, \eta) 
    +(-1)^{\lambda} f(x_2,a)G_{0}(k x_2, \eta)].
    \label{asymptotic-part}
\end{eqnarray}
The normalization of the asymptotic wavefunctions is chosen such that
\begin{equation}
    \mel{\Omega_1^{\sigma}}{H-E}{\Omega_2^{\sigma}}-\mel{\Omega_2^{\sigma}}{H-E}{\Omega_1^{\sigma}}=i.
    \label{normalisation-of-asympt-wf}
\end{equation}
The functions $F_{0}$ and $G_{0}$ are the regular and irregular Coulomb wavefunctions for a zero angular momentum, $R_{10}$ is a hydrogenic function with reduced mass $\mu_{12}=\alpha m_1$, principal quantum number $n_1=1$ and zero angular momentum. In addition, one has
\begin{align}
    \dfrac{1}{\mu_{12,3}} &=\dfrac{1}{m_1+m_2}+\dfrac{1}{m_3},\\
    k &= \sqrt{{2\mu_{12,3} \left(E-E_{n_1=1} \right)}},\\ 
    \text{and} \ \ \eta  &=\dfrac{(Z_1+Z_2)Z_3\mu_{12,3}}{k},
\end{align}
where $E$ is the total energy of the three-body system. The function $f(x_2,a)=1-e^{-a x_2}$, which depends on the parameter $a$, enforces the right behaviour at the origin on $G_{0}$. In practice, $a=k$ is chosen in this work. In the case of the scattering of an electron onto a hydrogen atom, $\eta=0$, and the Coulomb functions correspond to sine and cosine functions, respectively. 

The complex Kohn variational principle \cite{Kohn48,MJ87_CKVM,Lucchese89,Kievsy97,AMO23_Drake} states that the $\mathcal{S}$ matrix at the energy $E$ or, equivalently, wavevector $k$, can be evaluated by
\begin{equation}
    \mathcal{S}(k)={\Bar{\mathcal{S}}}(k)+i \mel{\Psi^{\sigma}}{H-E}{\Psi^{\sigma}}.
    \label{SecondOrderKohn}
\end{equation}
The estimate $\Bar{\mathcal{S}}(k)$ of the $\mathcal{S}$ matrix is determined by imposing that the Schrödinger equation is fulfilled on the subspace of dimension $N_t+1$ spanned by the functions $\{{\phi}_{l}^{\sigma},\Omega_2^{\sigma}\}$, i.e.
\begin{equation}
    \begin{cases}
        \mel{{\phi}_{l}^{\sigma}}{H-E}{{\Psi}^{\sigma}}&=0\\
        \mel{\Omega_2^{\sigma}}{H-E}{{\Psi}^{\sigma}}&=0
    \end{cases}
    \label{EqsKohn}
\end{equation}
for $l=1,...,N_t$. In the following, we use the notations
\begin{align}
    (\mathcal{H}_E)_{lm}&= \mel{\phi_{l}^{\sigma}}{H-E}{\phi_{m}^{\sigma}}\label{M0}\\
  {\rm{and  }\ \ \  } (\bm{\omega}^{\sigma}_{\lambda})_{l} &=\mel{\phi_{l}^{\sigma}}{H-E}{\Omega_{\lambda}^{\sigma}},\label{wlambda}
    \end{align}
where $l,m=1,...,N_t$ and $\lambda=1,2$. By expressing the wavefunction $\Psi^{\sigma}$ as its square-integrable and asymptotic contributions, see Eqs.~\eqref{square-integrable-part}~and~\eqref{asymptotic-part} respectively, the system of equations~\eqref{EqsKohn} becomes
\begin{equation}
    \begin{cases}
        &\mathcal{H}_E \bm{C} +\bm{\omega}^{\sigma}_{1}+ \Bar{\mathcal{S}}(k) \bm{\omega}^{\sigma}_{2}=0\\
        &(\bm{\omega}^{\sigma}_{2})^T \bm{C}+\mel{\Omega^{\sigma}_{2}}{H-E}{\Omega_{1}^{\sigma}+\Bar{\mathcal{S}}(k)\Omega_{2}^{\sigma}}=0
    \end{cases},
\end{equation}
where $\bm{C}$ contains the expansion coefficients of Eq.~\eqref{square-integrable-part}. By solving this system, one obtains a first estimate of the $\mathcal{S}$ matrix,
\begin{equation}
    \Bar{\mathcal{S}}(k)=\dfrac{\mel{\Omega^{\sigma}_{2}}{H-E}{\Omega_{1}^{\sigma}}-(\bm{\omega}^{\sigma}_2)^T\mathcal{H}_E^{-1} \bm{\omega}^{\sigma}_1}{\mel{\Omega^{\sigma}_{2}}{H-E}{\Omega_{2}^{\sigma}}-(\bm{\omega}^{\sigma}_2)^T \mathcal{H}_E^{-1} \bm{\omega}^{\sigma}_2}.
    \label{SfirstOrder}
\end{equation}
 By injecting Eq.~\eqref{SfirstOrder} in Eq.~\eqref{SecondOrderKohn}, an improved estimation of the $\mathcal{S}$ matrix, noted $\mathcal{S}(k)$, is obtained. The elastic phase shift $\delta_0(k)$ is directly deduced from $\mathcal{S}(k)$ by 
\begin{equation}
    \tan \delta_0(k) = i\dfrac{1-\mathcal{S}(k)}{1+\mathcal{S}(k)}.
\end{equation}
\subsection{The Lagrange-mesh method}
\label{LMMsection}
Following Refs.~\cite{HB99,DES22}, the $S$-state square-integrable wavefunction $\Phi^{\sigma}$ is expressed in perimetric coordinates, which are the set of three radial coordinates $(x,y,z)$ depending on the interparticle distances \cite{Pe58},
\begin{equation}
    \begin{cases}
    x&= r_{12}-r_{23}+r_{13}\\y&= r_{12}+r_{23}-r_{13}\\z&= -r_{12}+r_{23}+r_{13}
    \end{cases}.
    \label{perim_coord_def}
\end{equation}
The square-integrable wavefunction is expanded in a similar way as in bound-state calculations \cite{HB99},
\begin{eqnarray}
    \Phi^{\sigma}(x,y,z)= \dfrac{1}{2} \sum_{p=1}^{N_x} \sum_{q=1}^{N} \sum_{r=1}^{q-\sigma} C_{pqr} [&&F_{pqr}(x,y,z)\nonumber \\+(-1)^{\sigma} &&F_{pqr}(x,z,y)],
    \label{WF_expansion}
\end{eqnarray}
where the functions $F_{pqr}(x,y,z)$ read
\begin{equation}
    F_{pqr}(x,y,z)= \dfrac{f_p^{(N_x)} \left(x/h_x\right)f_q^{(N)} \left(y/h\right) f_r^{(N)} \left(z/h\right)}{\mathcal{N}_{pqr}(1+\delta_{qr})^{1/2}},
\end{equation}
and
\begin{equation}
    \mathcal{N}_{pqr}=\dfrac{\pi}{2} \sqrt{h_x h^2 (h_x x_p+h y_q)(h_x x_p +h z_r)(h y_q + h z_r)}.
\end{equation}
The quantities $h_x$ and $h$ are scale parameters. The permutation symmetry with respect to the electrons imposes that the number $N$ of basis functions in the variables $y$ and $z$ is the same, as well as their scale parameter $h$. As can be seen from Eq.~\eqref{perim_coord_def}, the effect of $P_{23}$ is indeed to exchange the variables $y$ and $z$. The total number of basis functions is given by $N_t = N_x N [N+(-1)^{\sigma}]/2$. \\ \indent
The Lagrange-Laguerre basis functions $\{f^{(\nu)}_{l} (\xi)\}_{l=1,...,\nu}$ are explicitly defined in Eq.~(19) of Ref.~\cite{DES22}, for instance. They consist of a product between a polynomial of degree $\nu-1$ and an exponential. They verify the important property 
\begin{equation}
    f^{(\nu)}_{l} (\xi_m) = \dfrac{\delta_{lm}}{\sqrt{\lambda_{m}}},
\end{equation}
where the $\{\xi_m\}_{m=1,...,\nu}$ are the zeros of the Laguerre polynomial of degree $\nu$ corresponding to the abscissae of the Gauss quadrature associated to the mesh, while $\{\lambda_m\}_{m=1,...,\nu}$ are the corresponding weights. The Lagrange-mesh method (LMM, see Ref.~\cite{Ba15}) consists of evaluating approximately all Hamiltonian matrix elements by means of the Gaussian quadrature associated to the three-dimension Lagrange-Laguerre mesh $\{(h_x x_p,h y_q,h z_r)\}_{p=1,...,N_x;\ q=1,...,N;\ r=1,...,N}$. It leads to a diagonal potential matrix and a sparse Hamiltonian matrix. In addition, the basis is orthonormal at the Gauss approximation. Despite its simplicity and the approximate treatment of the matrix elements, the LMM is often as accurate as a purely variational method. A wavefunction similar to $\Phi^{\sigma}(x,y,z)$ has been successfully used to study the $S$-bound states of diverse three-body systems (He, Ps$^-$, H$_2^+$, see Ref.~\cite{HB99}), as well as the resonance states of He and Ps$^-$, by combining the LMM and the complex scaling method \cite{DES22}.  \\ \indent
The element of volume in perimetric coordinates reads, after integrating over the angles,
\begin{equation}
    \text{d}V = \dfrac{\pi^2}{4} (x+y) (x+z) (y+z)\text{d}x \text{d}y \text{d}z.
\end{equation}
The link between the radial Jacobi coordinates and the perimetric ones is given by
\begin{equation}
    x_1=\dfrac{x+y}{2}
\end{equation}
and
\begin{equation}
    x_2=\dfrac{1}{2}\sqrt{\alpha^2 (x+y)^2+2 \alpha [yz-x(x+y+z)]+(x+z)^2}.
    \label{LinkPerimJac}
\end{equation}

\subsection{Resonance parameters}
\label{resonanceParameters}
A resonance corresponds to a pole of the $\mathcal{S}$ matrix. To determine this pole, we follow a procedure described in Ref.~\cite{RA07}, which consists of approximating the $\mathcal{S}$ matrix as a rational function of the type
\begin{equation}
    \mathcal{S}(k) = \dfrac{1+\sum_{n=1}^{N_P}a_n k^n}{1+\sum_{n=1}^{N_P}(-1)^n a_n k^n}.
    \label{Sapprox}
\end{equation}
The complex coefficients $a_n$ are obtained by equalizing the $\mathcal{S}$ matrix computed by the Kohn variational principle and its approximation \eqref{Sapprox}. The poles of the $\mathcal{S}$ matrix correspond to the roots of the denominator of Eq.~\eqref{Sapprox}. A pole corresponding to a resonance remains stable with respect to an increase of the number of interpolation points $N_P$. By noting this pole $k_{\rm{res}}$, the resonance energy $E_r$ and width $\Gamma$ are then deduced from
\begin{equation}
    E_r-i\dfrac{\Gamma}{2}=\dfrac{k_{\rm{res}}^2}{2 \mu_{12,3}}+E_{n_1=1}.    
\end{equation}
%--------------------------------------------------------------------
% RESULTS
%--------------------------------------------------------------------
\section{Results\label{results}}

In this section, we present the phase shifts and the resonance parameters related to the elastic scattering of an electron onto a hydrogen atom ($Z_1=1$). We consider the case in which the proton is of infinite mass ($m_1\rightarrow \infty$, $\alpha=0$), which we note $^{\infty}$H$^-$, as well as the case in which $m_1=1836.152 673 43$ (2018 CODATA recommended value \cite{CODATA18}). 

\subsection{Methodology}
The computation of the Hamiltonian matrix using the Lagrange-mesh method is rather simple and described in Refs.~\cite{HB99,HB01}. Considering a finite proton mass does not add any difficulty to the treatment of the Hamiltonian matrix elements of Eq.~\eqref{M0}.\\ \indent
The computation of the hybrid vectors of Eq.~\eqref{wlambda} is performed in perimetric coordinates. On one hand, in the infinite proton mass case, a three-dimensional Gauss-Laguerre quadrature is used. The number of points used for the quadratures related to each radial perimetric coordinates is chosen as about twice the number of basis functions of each coordinate (typically, 30 to 80 points). The components of the hybrid vectors have an absolute accuracy of 13 to 14 digits. On the other hand, in the finite proton mass case, a combination of Gauss-Legendre and Gauss-Laguerre quadratures is needed to reach a satisfying absolute accuracy, of about 7 significant digits. The number of points used to integrate in each of the three dimensions is comprised between 100 and 250. This important difference of treatment may be explained by considering the variable $x_2$, which is a simple linear combination of the perimetric coordinates in the former case (namely, $x_2=r_{13}=(x+z)/2$ for $^{\infty}$H$^-$), whilst it is much more complicated in the latter, see Eq.~\eqref{LinkPerimJac}. \\ \indent
The matrix elements between asymptotic functions are computed in Jacobi coordinates. In the infinite proton mass case, all asymptotic matrix elements are analytical. In the finite proton mass case, the computation of the unpermuted asymptotic matrix elements remains analytical, but the permuted ones need a numerical evaluation since the permuted Jacobi coordinates have a complicated expression:
\begin{align}
    &P_{23}x_1 = \sqrt{x_2^2+\alpha^2x_1^2+2\alpha x_1 x_2 u},\\
    &P_{23}x_2 = \sqrt{\alpha^2 x_2^2+(1-\alpha^2)^2 x_1^2+2 \alpha (\alpha^2-1) x_1x_2 u},
\end{align}
where $u={\bm{x}_1 \cdot \bm{x}_2}/{(x_1 x_2)}$. A multipolar expansion of the Coulomb potential is performed and the resulting integrals are evaluated by means of Gaussian quadratures. This approach leads to an absolute accuracy of at least 10 significant digits on the asymptotic matrix elements. A useful numerical check is provided by the identity
\begin{equation}
    \mel{\Tilde{\Omega}_1}{H-E}{\Omega_2}-\mel{\Tilde{\Omega}_2}{H-E}{\Omega_1}=0.
\end{equation}
Finally, to solve the system of equations \eqref{EqsKohn}, we use the package ILUPACK \cite{ILUPACK}, which is suited for large symmetric sparse systems. The unitarity of the $\mathcal{S}$ matrix is used as a criterion of convergence, as illustrated in Table~\ref{convPS}.

\subsection{Phase shifts \label{PS_section}}
\begin{figure}
\includegraphics[scale=0.7]{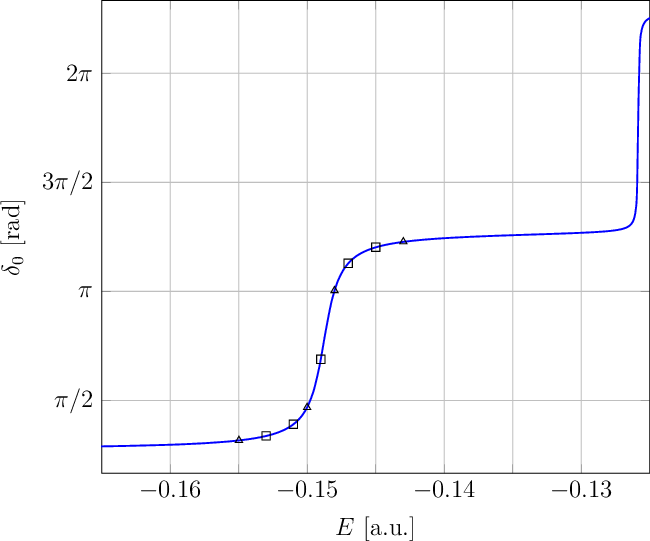}
\caption{\label{PS_1S} $S$-wave phase shift in the singlet case of $e+^{\infty}$H scattering as a function of the total energy $E$, below the $n_1=2$ ionization threshold. The points used for extracting the resonance parameters are highlighted (see the discussion in the text). }
\end{figure}

Table~\ref{convPS} presents the convergence of the singlet and triplet phase shifts of $^{\infty}$H$^-$ with respect to an increase of the size of the mesh, for $k=0.2$. The deviation from unitarity of the $\mathcal{S}$ matrix corresponding to each mesh size is displayed. The values of the scale parameters $h$ and $h_z$ as well as the regularization coefficient $a$ can be varied by about 10\% without affecting the converged results, which are accurate up to a change of two units on the last displayed digit. The largest considered mesh size is around $N_x=15$ and $N=30-35$. For this size of the mesh, the calculation of the phase shift takes a few minutes on a standard workstation.  \\ \indent
Table~\ref{PhaseShifts} presents the converged singlet and triplet phase shifts of $^{\infty}$H$^-$, for values of $k$ ranging from $0.1$ to $0.8$ by step of $0.1$. These phase shifts are accurate up to a change of a few units on the last displayed digit. {In all cases, the convergence rate is similar to the one presented in Table~\ref{convPS}. The converged value is reached around $(N_x,N)=(10,35)$, and the results are stable up to a change of about 10\% of the scale parameters or of the regularization coefficient $a$. However, one should note that for the same mesh size, the unitarity of the $\mathcal{S}$ matrix is poorer as the energy gets closer to the $n_1=2$ ionization threshold. The two closed channels corresponding to the $n_1=2$ excited states of hydrogen become indeed more and more important as the energy increases. By including explicitly these two closed channels in the asymptotic part of the wavefunction, the unitarity of the $\mathcal{S}$ matrix could be probably enhanced close to the $n_1=2$ ionization threshold.} \\ \indent
A comparison with results obtained in 2001 by Bhatia and Temkin \cite{BT01_CKVM} is provided {in Table~\ref{PhaseShifts}}. Their results being accurate up to a change of two units on the fourth digit after the decimal point, our results agree in most cases, except in the singlet case at $k=0.3$, remarkably. We believe there is a small typo in this case, in Ref.~\cite{BT01_CKVM}. Indeed, less accurate results have been obtained by Schwartz in 1961, in his pioneering work about the electron-hydrogen scattering \cite{Schwartz61}, and the singlet-state result from Ref.~\cite{BT01_CKVM} at $k=0.3$ does not agree with Schwartz's, contrary to our results, which agree with Schwartz's in all cases. {For $k=0.2$, our results are compared to the more accurate ones obtained in Ref.~\cite{ZMV08}, and the agreement is excellent.}\\ \indent
Figure~\ref{PS_1S} illustrates the singlet-state phase shift computed for an energy comprised between $E=-0.165$ ($k\approx 0.8185$) and the $n_1=2$ ionization threshold ($k \approx 0.866$). In the singlet case, there are two resonances below this threshold. This figure is very similar to Figure 1 of Ref.~\cite{KH96}. The two resonances clearly appear, the first one being wider than the second one. The precise values of the resonance parameters are presented in the next subsection.  \\ \indent
Finally, Table~\ref{PhaseShifts_fin} gathers the results for the hydrogen ion with a finite proton mass. The mesh parameters and the regularization coefficients are the same as for the infinite proton mass case. Compared to this latter case, the phase shift is systematically smaller. We could not compare our results with other works accurate enough to assess the sensitivity of the results with respect to the inclusion of a finite proton mass. {To our knowledge, the only phase shifts computed for a finite proton mass are reported in Ref.~\cite{KH96}, in which the tangents of the phase shifts are accurate up to a 1\% error. In order to provide a comparison with their results, we have considered additional values of the wavevector, close to the resonance energies, in Table~\ref{PhaseShifts_fin}. Our results are about three orders of magnitude more accurate.} \\ \indent
In conclusion, the computed phase shifts have an accuracy similar to the best one reported in literature in the case of an infinite proton mass, whilst it is greater by {up to three} orders of magnitude in the finite proton mass case. This shows the high accuracy obtained by combining the Lagrange-mesh method and the complex Kohn variational method to compute the $\mathcal{S}$ matrix. 

\begin{table}
  \caption{\label{convPS} Convergence of the phase shift [rad] as a function of the size of the Lagrange mesh, for $^{\infty}$H, at the energy $E=-0.48$ ($k=0.2$). The scale parameters are $(h_x,h)=(1.0,1.3)$ for the singlet case and $(1.2,1.5)$ for the triplet case. The regularization coefficient is $a = 0.2$. Atomic units are used.}
  \begin{ruledtabular}
\begin{tabular}{rrl|S[table-format = 2.8]S[table-format = 1e+2]|S[table-format = 3.8]S[table-format = 1e+2]}
        & & & \multicolumn{2}{c}{{$^1S$}} & \multicolumn{2}{c}{{$^3S$}}\\
    $N_x$ & $N$ &  & {$\delta_0(k)$} & {$\abs{1-\abs{\mathcal{S}}^2}$} & {$\delta_0(k)$} & {$\abs{1-\abs{\mathcal{S}}^2}$}\\
\hline \rule{0pt}{1.\normalbaselineskip}
    10 & 20  & &  2.0669710 & 3E-08 & -0.4240955 & 2E-08\\
    10 & 25  & &  2.0669867 & 5E-09 & -0.4240879 & 8E-09\\
    10 & 30  & &  2.0669921 & 3E-10 & -0.4240845 & 4E-09\\
    10 & 35  & &  2.0669942 & 2E-10 & -0.4240827 & 3E-09\\
    15 & 35  & &  2.0669941 & 2E-10 & -0.4240827 & 3E-09\\
    \multicolumn{2}{l}{Converged} & & 2.06699 &  & -0.42408 & \\
\end{tabular}
\end{ruledtabular}
\end{table}

\begin{table}    \caption{\label{PhaseShifts}Singlet and triplet phase shifts [rad] for $(N_x,N)=(10,35)$ of $^{\infty}$H$^-$, for different values of the wave vector $k$ ($a=k$). Results are accurate up to a change of a few units on the last displayed digit. The error on the results of Ref.~\cite{BT01_CKVM} is up to two units on the fourth digit after the decimal point. Atomic units are used.}
\begin{ruledtabular} 
\begin{tabular}{lrr|ccS[table-format = 2.6]|ccS[table-format = 3.6]}
    & &  & \multicolumn{3}{c}{{$^1S$}} & \multicolumn{3}{c}{{$^3S$}}\\
    $k$ & $E$ & & $h_x$ & $h$ & {$\delta_0(k)$} & $h_x$ & $h$ & {$\delta_0(k)$}\\   \hline \rule{0pt}{1.\normalbaselineskip}0.1 & $-$0.495 & & 1.0 & 1.3 & 2.55374 & 1.2 & 1.5 & -0.20303\\
    \multicolumn{2}{r}{Ref.~\cite{BT01_CKVM}} & & & & 2.55358 & & & -0.20306\\
    0.2 & $-$0.480 & & 1.0 & 1.3 & 2.06699 & 1.2 & 1.5 & -0.42408\\
    \multicolumn{2}{r}{Ref.~\cite{ZMV08}} & & & & 2.06699 & & & -0.42409\\
    \multicolumn{2}{r}{Ref.~\cite{BT01_CKVM}} & & & & 2.06678 & & & -0.42418\\
    0.3 & $-$0.455 & & 1.0 & 1.3 & 1.69684 & 1.2 & 1.5 & -0.64172\\    
    \multicolumn{2}{r}{Ref.~\cite{BT01_CKVM}} & & & & 1.69816 & & & -0.64184\\
    0.4 & $-$0.420 & & 1.0 & 1.4 & 1.41557 & 1.2 & 1.6 & -0.84735\\
    \multicolumn{2}{r}{Ref.~\cite{BT01_CKVM}} & & & & 1.41540 & & &-0.84751\\
    0.5 & $-$0.375 & & 1.0 & 1.4 & 1.20109 & 1.2 & 1.6 & -1.03683\\
    \multicolumn{2}{r}{Ref.~\cite{BT01_CKVM}} & & & & 1.20094 & & & -1.03705\\
    0.6 & $-$0.320 & & 1.0 & 1.4 & 1.04113 & 1.2 & 1.6 & -1.20839\\
    \multicolumn{2}{r}{Ref.~\cite{BT01_CKVM}} & & & & 1.04083 & & & -1.20887\\
    0.7 & $-$0.255 & & 1.2 & 1.4 & 0.93098 & 1.4 & 1.7 & -1.36169\\
    \multicolumn{2}{r}{Ref.~\cite{BT01_CKVM}} & & & & 0.93111 & & & -1.36209\\
    0.8 & $-$0.180 & & 1.2 & 1.4 & 0.88773 & 1.4 & 1.7 & -1.49725\\
    \multicolumn{2}{r}{Ref.~\cite{BT01_CKVM}} & & & & 0.88718 & & &-1.49780\\
\end{tabular}
\end{ruledtabular}
\end{table}

\begin{table}        
\caption{\label{PhaseShifts_fin}Singlet and triplet phase shifts [rad] for $(N_x,N)=(10,35)$ of H$^-$, for different values of the wavevector $k$ ($a=k$, $m_p=1836.152 673 43 $). Results are accurate up to a change of a few units on the last displayed digit. The results of Ref.~\cite{KH96} are accurate up to 1\%. Atomic units are used.}
\begin{ruledtabular} 
\begin{tabular}{l|ccS[table-format = 2.6]|ccS[table-format = 3.6]}
     & \multicolumn{3}{c}{{$^1S$}} & \multicolumn{3}{c}{{$^3S$}}\\
    $k$ & $h_x$ & $h$ & {$\delta_0(k)$} & $h_x$ & $h$ & {$\delta_0(k)$}\\     \hline \rule{0pt}{1.\normalbaselineskip}0.1 & 1.0 & 1.3 & 2.55329 & 1.2 & 1.5 & -0.20318\\
    0.2 & 1.0 & 1.3 & 2.06631 & 1.2 & 1.5 & -0.42438\\
    0.3 & 1.0 & 1.3 & 1.69608 & 1.2 & 1.5 & -0.64215\\
    0.4 & 1.0 & 1.4 & 1.41479 & 1.2 & 1.6 & -0.84787\\
    0.5 & 1.0 & 1.4 & 1.20034 & 1.2 & 1.6 & -1.03743\\
    0.6 & 1.0 & 1.4 & 1.04045 & 1.2 & 1.6 & -1.20904\\
    0.7 & 1.2 & 1.4 & 0.93042 & 1.4 & 1.7 & -1.36237\\
    0.8 & 1.2 & 1.4 & 0.88765 & 1.4 & 1.7 & -1.49793\\
    0.8325 & 1.2 & 1.4 & 1.05453 & 1.4 & 1.7 & -1.53828\\
    {Ref. \cite{KH96}$\, $}& & & 1.06 & & & {$-$}\\ 
    0.8366 & 1.2 & 1.4 & 1.62750 & 1.4 & 1.7 & -1.54324\\
    {Ref. \cite{KH96}$\, $}& & & 1.643 & & & {$-$}\\
%    0.8631 & 1.2 & 1.4 & 4.0363 & 1.4 & 1.7 & -1.5575\\
%    {Ref.\cite{KH96}$\, $}& & & {$-$} & & & -1.545\\
\end{tabular}   
\end{ruledtabular}
\end{table}

\subsection{Resonances}
In this section, we extract the resonance parameters from values of the $\mathcal{S}$ matrix computed at $N_P$ real energies. We check the convergence of the resonance parameters with respect to an increase of the number of extrapolation points $N_P$, as well as to an increase of the size of the Lagrange mesh. The sensitivity of the results with respect to a change of 5 to 10\% of the scale parameters and of the regularization coefficient is also checked. \\ \indent
The procedure for checking the convergence with respect to $N_P$ is as follows. We start with an interval of energy roughly centred around the studied resonance energy, and having the extension of a few times the width. We compute the $\mathcal{S}$ matrix at $N_P$ equally spaced points belonging to the chosen real-energy interval. By using the method shortly described in Sec.~\ref{resonanceParameters}, the resonance parameters are determined. We add extrapolation points inside the interval as well as outside it until the convergence is reached.\\ \indent
An example of such a convergence study is presented in Table~\ref{convRes} for the first singlet resonance state of $^{\infty}$H$^-$, which appears in Fig.~\ref{PS_1S}. The five extrapolation points (square marks) used for obtaining the first line of Table~\ref{convRes}, as well as the four additional extrapolation points (triangle marks) used to obtain the third line of Table~\ref{convRes}, are highlighted in Fig.~\ref{PS_1S}. The obtained results reach an absolute accuracy of about $10^{-10}$ for the first resonance, for which the initial search interval is $[-0.153,-0.145]$. The regularization parameter $a$ is equal to the mean wavevector along the chosen interval.\\ \indent
Table~\ref{ResonancesTable} gathers converged results obtained for the resonances of the singlet and triplet states of the hydrogen ion, below the $n_1=2$ threshold. The finite and infinite proton mass cases are considered. The mesh parameters are reported in the Table. They are the same in both finite and infinite proton mass cases, and the reached accuracies are of the same order of magnitude. The absolute accuracy is about $10^{-8}$ for the $^1S(2)$ resonance, and $10^{-11}$ for the $^3S(1)$ resonance.\\ \indent
We provide a comparison of these resonance parameters with results obtained by applying the complex scaling method (CSM) on a Lagrange mesh, following Ref.~\cite{DES22}.  {The mesh parameters used in the complex scaling method are the same as in the Kohn method, except for the second singlet-state resonance, for which $(h_x,h)=(1.8,3.5)$.} The CSM results have an absolute accuracy {being higher by two to three orders of magnitude compared to the Kohn variational principle results}, i.e. $10^{-13}$, $10^{-11}$ and $10^{-13}$ for the $^1S(1)$, $^1S(2)$ and $^3S(1)$ resonances, respectively. There is a 10-digit agreement between the results obtained with the two approaches. An additional comparison with results from the literature, reaching an absolute accuracy up to $10^{-6}$, is also provided (see Refs.~\cite{KH96,CH07,DGH23}). Note that the results displayed in Ref.~\cite{KH96} have been computed with a slightly different proton mass ($m_p=1836.1515$) from ours, but this difference is irrelevant {at their level of accuracy}. In all cases, our results are more accurate by several orders of magnitude. {Note however that in the singlet case for both our methods, there is a quite important loss of accuracy between the $^1S(1)$ and $^1S(2)$ resonance parameters. We believe that by including explicitly the aforementioned closed channels in the asymptotic expansion (see Sec.~\ref{PS_section}), one could improve the accuracy of the $^1S(2)$ resonance parameters.}
\begin{table}
     \caption{\label{convRes} Convergence of the resonance parameters [a.u.] for the first singlet-state resonance of $^{\infty}$H$^-$, as a function of the number of extrapolation points $N_P$ as well as the size of the mesh. The scale parameters are $(h_x,h)=(1.2,1.4$).}
\begin{ruledtabular}
\begin{tabular}{lccS[table-format = 2.11]S[table-format = 1.7e+2]}
     $N_P$ & $N_x$ & $N$ & {$E_r$} & {$\Gamma$}\\ 
    \hline \rule{0pt}{1.\normalbaselineskip}5 & 10 & 25 &  -0.14877615703& 1.7336721E-03\\
    7 & 10 & 25 &  -0.14877625498& 1.7332404E-03\\
    9 & 10 & 25 &  -0.14877625497& 1.7332405E-03\\
    11& 10 & 25 &  -0.14877625497& 1.7332405E-03\\
      & 15 & 25 &  -0.14877625581& 1.7332398E-03\\
      & 20 & 25 &  -0.14877625582& 1.7332398E-03\\
    & 20 & 35 & -0.14877625345 & 1.7332353E-03\\
      & 20 & 40 & -0.14877625360 & 1.7332369E-03\\
       & 20 & 45 & -0.14877625413 & 1.7332368E-03\\
     \multicolumn{3}{l}{Converged} & -0.148776254 & 1.733237E-03 \\
\end{tabular}     
\end{ruledtabular}
\end{table}
\begin{table*}
\caption{\label{ResonancesTable}Resonance parameters [a.u.] for the singlet and triplet resonant states of the hydrogen ion, with finite (H$^-$) and infinite ($^{\infty}$H$^-$) proton mass. A comparison with results obtained by applying the complex scaling method (CSM), as well as results from the literature, is provided. The mesh parameters are given in the text. Results are accurate up to a change of a few units on the last displayed digit.}
\begin{ruledtabular}
\begin{tabular}{lrlllllS[table-format = 2.12]S[table-format = 1.9e+2]}
    & & &  {$N_x$} &  {$N$} &  {$h_x$} &  {$h$} &  {$E_r$} & {$\Gamma$}\\ 
    \hline \rule{0pt}{1.\normalbaselineskip}
    $^1S(1)$ & H$^-$ & Present & 15 & 40 & 1.2 & 1.4 &  -0.148694751 & 1.730756E-03 \\
    & & CSM & 15 & 40 & 1.2 & 1.4 & -0.148694751429 & 1.730755239E-03\\
    & & \multicolumn{1}{l}{Ref.~\cite{KH96}} & & & & & -0.1487 & 1.735E-03 \\
    & $^{\infty}$H$^-$ & Present & 15 & 40 & 1.2 & 1.4  & -0.148776254 & 1.733237E-03 \\
    & & CSM & 15 & 40 & 1.2 & 1.4 & -0.148776253939 & 1.733236366E-03 \\
    & & \multicolumn{1}{l}{Ref.~\cite{CH07}} & & & & & -0.148776 & 1.732E-03 \\
    & & \multicolumn{1}{l}{Ref.~\cite{DGH23}} & & & & & -0.14877 & 1.735E-03 \\
    & & \multicolumn{1}{l}{Ref.~\cite{KH96}} & & & & &  -0.1488 & 1.735E-03 \\
    \hline \rule{0pt}{1.\normalbaselineskip}
    $^1S(2)$ & H$^-$ & Present & 15 & 45 & 1.4 & 2.2 & -0.1259514 & 9.03E-05 \\
    & & CSM & 15 & 45 & 1.8 & 3.5 &  -0.1259514447 &   9.04284E-05\\
    & & \multicolumn{1}{l}{Ref.~\cite{KH96}} & & & & & -0.12595 & 9.0E-05 \\
    & $^{\infty}$H$^-$ & Present & 15 & 45 & 1.4 & 2.2 & -0.1260201 & 9.06E-05 \\
    & & CSM & 15 & 45 & 1.8 & 3.5 &  -0.126 020 063 7 &   9.05298E-05\\
    & & \multicolumn{1}{l}{Ref.~\cite{KH96}} & & & & & -0.12604 & 9.1E-05 \\
    & & \multicolumn{1}{l}{Ref.~\cite{DGH23}} & & & & & -0.126 & {$-$} \\ \hline \rule{0pt}{1.\normalbaselineskip}
    $^3S(1)$ & H$^-$ & Present & 15 & 40 & 1.2 & 3.0 & -0.1270342774 & 6.816E-07 \\
    & & CSM & 15 & 40 & 1.2 & 3.0 &  -0.127034277368 & 6.81613E-07\\
    & & \multicolumn{1}{l}{Ref.~\cite{KH96}} & & & & & -0.12705 & 5.29E-07 \\
    &  $^{\infty}$H$^-$ & Present & 15 & 40 & 1.2 & 3.0 &  -0.1271042116 & 6.843E-07 \\
    & & CSM & 15 & 40 & 1.2 & 3.0 & -0.127104211642 & 6.84271E-07\\
    & & \multicolumn{1}{l}{Ref.~\cite{KH96}} & & & & & -0.12710 & 5.29E-07 \\
\end{tabular} 
\end{ruledtabular}
\end{table*}
\section{Conclusion\label{conclu}}
In this paper, we showed that combining the Lagrange-mesh method in perimetric coordinates and the complex Kohn variational method is appropriate for computing accurately the $\mathcal{S}$ matrix describing an elastic scattering in three-body Coulomb systems. We illustrated this approach by considering the $S$-wave scattering of an electron onto a hydrogen atom, between the $n_1=1$ and $n_1=2$ ionization thresholds. \\ \indent
On one hand, we showed that the obtained elastic phase shifts are as accurate as the most accurate results presented in the literature with 6 significant digits, when considering an infinite proton mass. While most studies are restricted to the infinite proton mass case, the finite proton mass one is considered here as well, and the obtained phase shifts are as accurate as in the infinite mass case.\\ \indent
On the other hand, the evaluation of the $\mathcal{S}$ matrix at real energies can be used to determine its complex poles, and hence the resonance parameters associated to the scattering. The obtained resonance parameters have an absolute accuracy comprised between $10^{-8}$ and $10^{-11}$. This accuracy is greater than the one presented in the literature by several orders of magnitude. \\ \indent
{As future prospects, the investigation of resonances in exotic helium-like atoms could be conducted by means of the current approach. In particular, antiprotonic helium is currently under study \cite{SDE24b}}.

\section*{Acknowledgements}
The authors thank Alejandro Kievsky for enlightening discussions on the complex version of the Kohn variational principle, as well as Daniel Baye for his comments on the present manuscript. This work has received funding from the Fonds de la Recherche Scientifique-FNRS under Grant Nos. 4.45.10.08. One of the authors (JS) is a Research Fellow of the F.R.S.-FNRS.\\

\bibliography{references}
\end{document}